# THE DYNAMICS OF PUBLISHING OF POSTS AND COMMENTS ON FACEBOOK (THE RUSSIAN SEGMENT, THE FIRST FIVE MONTHS OF 2013)

A. V. Makarenko

Constructive Cybernetics Research Group,
P.O. Box 560, Moscow, 101000, Russia

Institute of Control Sciences RAS,
ul. Profsoyuznaya 65, Moscow, 117997, Russia

e-mail: avm.science@mail.ru

**Abstract.** The article presents a study of some characteristics of post and comment publishing in the Russian segment of Facebook. A number of non-trivial results has been obtained. For example, a significant anomaly has been detected in the number of user accounts with the rate of publishing posts of approximately two posts per three days. The analysis has been carried out at the level of basic characteristics that are shared by most social media platforms. It makes possible a direct comparison of obtained results with data from other platforms. The article presents an approach to formalization and ordering of structural and informational elements on social media platforms. The approach is based on the representation of these structural elements in the form of a coherent hierarchy of container objects and their relations. This method allows to structure and analyze raw data from different social media platforms in a unified algorithmic design. The described approach is more formal, universal and constructive than other known approaches.

**Keywords:** social media, data representation scheme, Facebook, dynamics of publishing, posts, comments.

## 1. Introduction

There is no doubt about the high degree of informatization of today's society. Its scope is so vast that a significant proportion (both in content and in volume terms) of various data about individuals and social groups is available on social media and public repositories of digital data. Social media is a set of online technologies that allow users to communicate with each other [1, 2]. This communication can take many forms: users can share their views, experience and knowledge, can interact with each other making new contacts, and can share news, information, photos, videos, music and hyperlinks to various content. In this context, content manipulation functions on the ideological and technological base of Web 2.0 (user-created content) [3].

The described phenomenon gives rise to various social processes with far-reaching implications. These processes manifest in various aspects: financial, political, cultural, scientific, etc. Comprehensive research of social media is essential for understanding current events, making forecasts and keeping negative tendencies in check. At the initial stage, it requires studying social media behavior at the levels of individuals, social groups, nations and of humanity as a whole.

The present paper focuses on the study of some characteristics of information activity of social media user accounts. The object of the study is the Russian segment of Facebook in the first four months of 2013. The study was carried out at the level of posts and comments on these posts. Specific properties of Facebook (such as "Likes") were not included in order to facilitate comparison of obtained results with data from other social media platforms.

This article is organized as follows. Section 2 describes the methodology of analysis and the format of input data presentation. Section 3 presents the results of the analysis of the dynamics of publishing of posts and comments. Section 4 presents general conclusions.

## 2. Methodology of Analysis and Initial Data

The approach to studying the dynamics of publishing of posts and comments that we have chosen is quite universal. First, a model has been developed for structuring and comparing data from different social media platforms within the framework of a unified construction. Second, the analysis has been carried out at the level of basic characteristics that are shared by most social media platforms. These solutions make it possible to directly compare the obtained results to data from different platforms.

We have implemented formalization and ordering of informational elements of social media in the form of a hierarchy of containers (see Fig. 1a):

**E** – (enviroment);

**P** – (platform) – a single social media platform;

**A** – (account) – a message published from a specific account;

**M** – (message) – a message from a specific user account (posts and comments);

**B** – (block) – a logically complete section of a message.

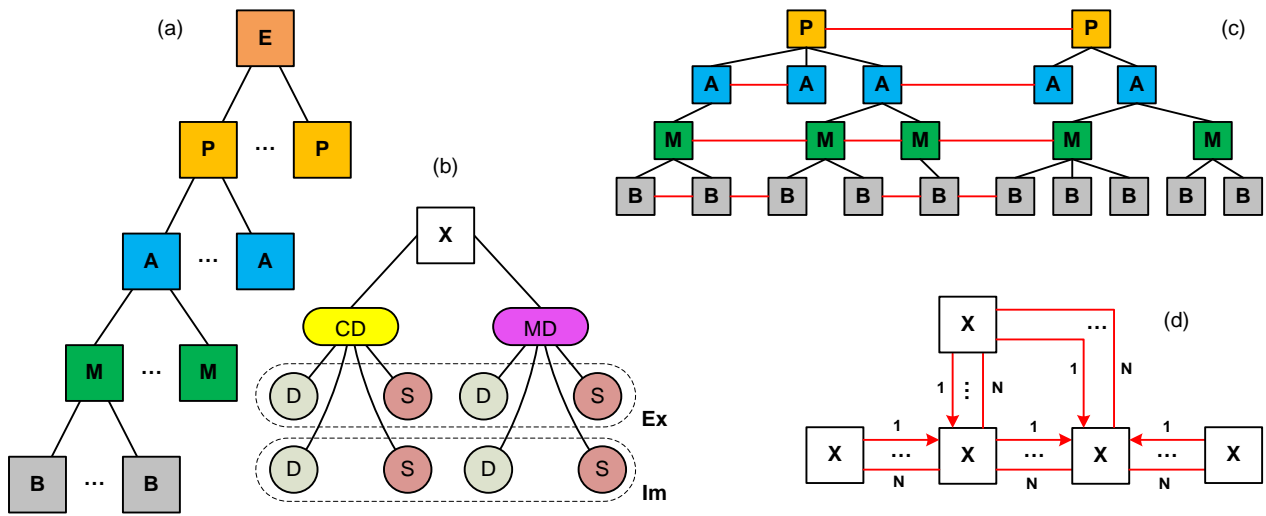

**Fig. 1.** Basic concepts of the approach to structuring and collation of data from different social media platforms: (a) – hierarchy of container objects; (b) – decomposition scheme of a container object; (c) – example of possible relations between objects; (d) – two types of relations (directional and non-directional).

Next, containers are decomposed according to the following scheme (see Fig. 1b). Each container consists of two constituents:

**CD** – (contents);

**MD** – (metadata).

Each of the constituents can contain the following components:

**D** – (data) – facts, information, indicators;

**S** – (sence) – sense, meaning, opinion.

Each of the components can be of two types:

**Ex** – (explicit) – explicitly given (at the level of formal semantic markup);

**Im** – (implicit) – implicitly given (inferred with varying level of confidence).

Apart from container objects and their constituents, connections between objects (relations of various kinds) are also an important structural component of social media and it is important to understand that these relations are established at the level of decomposition of objects (CD/MD, D/S). Usually these relations are:

- elementary – a relation has the one target identity;

- two-point – a relation where each elementary link unites two objects;

- horizontal – a relation that combines objects at the same level of the hierarchy, see Fig. 1c;

- homogeneous – a relation that unites identical constituents/components.

By fixing relations (their target identity) on the set of container objects (and/or their constituents/components), we obtain a graph of relations. It stands to mention that all the edges in the graph (at a fixed target identity) are either directional or non-directional (mixing is prohibited, see Fig. 1d). The graph can then be marked and weighed both by edges and by vertices, depending on the purpose of analysis and target identity of relations. Multiple edges (if any) are then replaced by a single edge, but the weights of multiple edges and degree of multiplicity are recorded as additional parameters into the weight of the final edge.

The approach to the representation of structural elements of social media as a hierarchy of container objects and their relations is more formal, universal and constructive than other known approaches [4, 5].

The hierarchy of objects and the relations between them are encoded as a tabular key-value store, as a relational database, or as RDF triples, depending on configuration of data and the purpose of analysis [6]. In particularly difficult cases, all three methods of data representation are usually combined (with regard to their ranges of adequacy). We implement full transition from character identifiers to variable size unsigned integer identifiers for encoding (from bit fields up to 64-bit number representation). It significantly speeds up the processing of data (including processing on GPU, clusters and GRID systems), allows to manage memory flexibly and to use standard tools for storage and processing. A detailed description of the encoding methods and of data processing is beyond the scope of this article and it is not included.

**Fig. 2.** of the representation of input data: (a) – posts; (b) – comments; (c) – reference tables.

For the purposes of this article, we have analyzed one social media platform (Facebook) and two levels of hierarchy (user accounts and messages). Messages have been divided into two types: posts and comments. The content of messages has not been analyzed. Two metadata parameters have been included into the analysis: author and date of publishing. The analysis has been limited to the posts published by the users on their own timelines (account pages). The initial data was provided by "Digital Society Laboratory (DSL)" under the agreement of cooperation with the Institute of Control Sciences RAS. These data have been processed and presented as arrays, as seen in Fig. 2a and 2b. These arrays are used as input data for the processing program. Numeric IDs have been associated with actual URLs by means of reference tables (see Fig. 2c).

## 3. Results and Discussion

Summary information on the analyzed period (for posts) is shown in Table 1.

**Table 1.**

Summary information on the analyzed period (for posts)

| Parameter | Representation | Value |
|---|---|---|
| Date of publication of the first analyzed post | $T_b^{pst}$ | January 1, 2013, 00:00:01 UTC |
| Date of publication of the last analyzed post | $T_e^{pst}$ | July 2, 2013, 01:10:01 UTC |
| Duration of the analyzed period | $T_{be}^{pst}$ | ~ 152 days |
| Number of published posts | $N^{pst}$ | 96 745 854 |
| Number of active user accounts | $N_{pst}^{uac}$ | 2 864 213 |

Table 1 can be used to determine basic *performance* (rate of publishing) for user accounts by the number of posts:

$$\tilde{S}_{uac}^{pst} = \frac{N^{pst}}{N_{pst}^{uac}} \approx 34 .$$

In fact, there is a strong variation in performance (4 orders of magnitude). Figure 3a shows the distribution $\delta_{pst}^{uac}$ of the number of user accounts by the number of published posts $S_{uac}^{pst}$. The median of $\bar{S}_{uac}^{pst} \mid P^{uac} = 9$, see Fig. 3b. Maximum performance of $\max S_{uac}^{pst} = 35922$.

In Figure 3a, we see a local maximum in dependence $\delta_{pst}^{uac}\left(S_{uac}^{pst}\right)$ in the area $S_{uac}^{pst} = 102$ (area f100). Detailed analysis indicates that the area of the anomaly f100, as a first-order approximation, has the following limits: $S_{uac}^{pst} \mid f100 = [85, 160]$, see Fig. 3c. They are detected by a deviation in the monotony and smoothness of dependence of $\delta_{pst}^{uac}\left(S_{uac}^{pst}\right)$. Excluding area f100, this dependence can be approximated with acceptable adequacy by the following formula:

$$\check{\delta}_{pst}^{uac}\left(S_{uac}^{pst}\right) = \frac{295376}{10^{-4}\left(S_{uac}^{pst}-1\right)^{2.78} + 0.1\left(S_{uac}^{pst}-1\right)^{1.545} + 1} .$$

Whereby the relative error of approximation does not exceed 0.137 on the interval of $1 \le S_{uac}^{pst} \le 245$. For $S_{uac}^{pst} > 245$ the error increases (due to the increased fluctuations of $\delta_{pst}^{uac}$). It is necessary to underscore that the interval of $1 \le S_{uac}^{pst} \le 245$ covers 98.22 % of all user accounts.

Figure 3d shows the difference:

$$\delta_{pst}^{\prime uac} = \delta_{pst}^{uac} - \check{\delta}_{pst}^{uac} .$$

As a first-order approximation, the summation of $\delta_{pst}^{\prime uac}$ value in area f100 gives us an estimate of $96623^{+0}_{-5181}$ for the total of abnormal user accounts. At the moment, we are performing the analysis to identify specific user accounts in area f100.

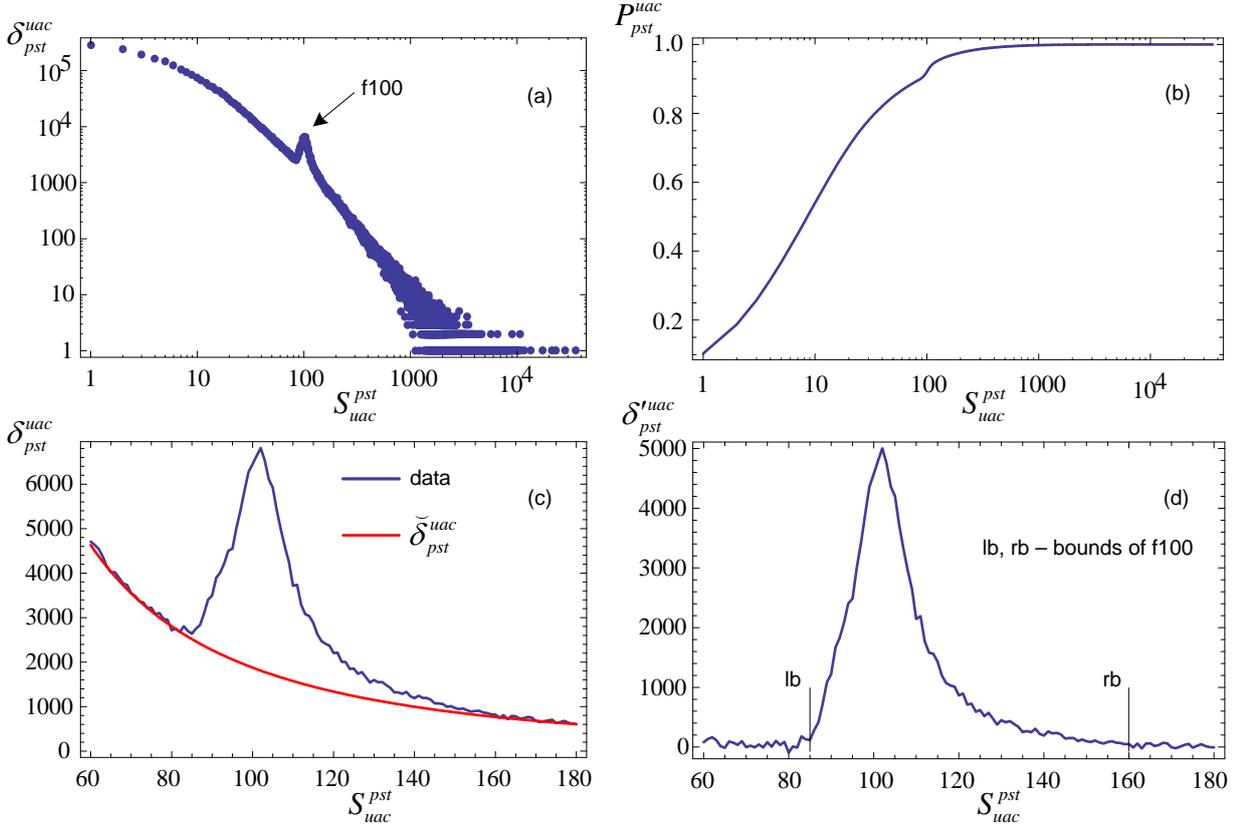

**Fig. 3.** (a) – dependence $\delta^{uac}_{pst}\left(S^{pst}_{uac}\right)$; (b) – cumulative share of user accounts, depending on the increase in performance $S^{pst}_{uac}$; (c) – a close up of the area of anomaly f100; (d) – graph of relative value $\delta^{ruac}_{pst}$ in area f100.

Share distribution of published posts $p^{pst}$ (of the total $N^{pst}$) as a function of the performance of user accounts $S^{pst}_{uac}$ is very informative. The distribution is defined by:

$$p^{pst}\left(S^{pst}_{uac}\right)=\frac{S^{pst}_{uac}\ \delta^{uac}_{pst}\left(S^{pst}_{uac}\right)}{N^{pst}}.$$

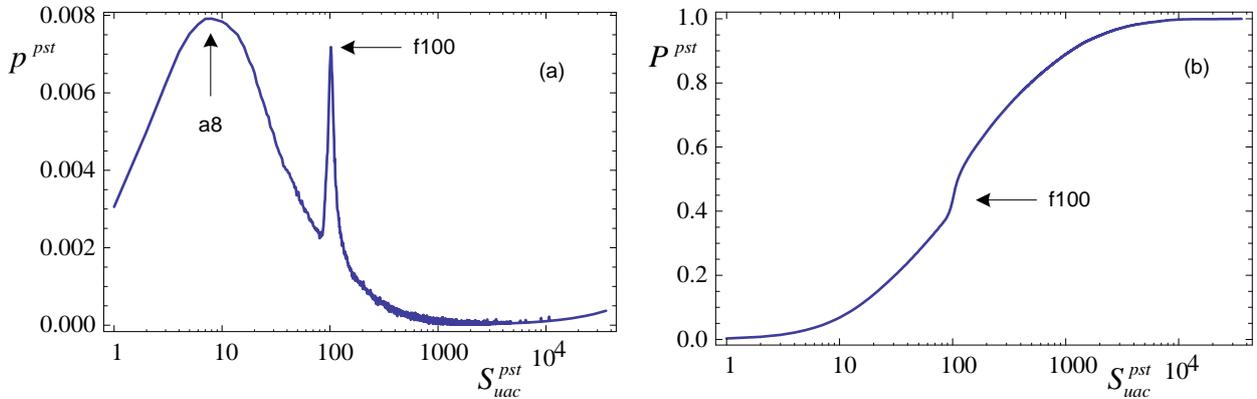

**Fig. 4.** (a) – Distribution of the share of posts $p^{pst}$ corresponding to the accounts with the performance of $S^{pst}_{uac}$; (b) – cumulative share of posts $P^{pst}$ in function $S^{pst}_{uac}$.

The distribution is shown in Fig. 4a where both areas f100 and a8, with the maximum of $S_{uac}^{pst}=8$, can be seen. In view of the discrete nature of the variables, the value of this maximum is consistent with the value of $\bar{S}_{uac}^{pst}=9$. Figure 4b shows that the median $\bar{S}_{uac}^{pst} \mid P^{pst}=112$.

Figure 5 is shows $\tau_{uac}^{pst}$, i.e. the distribution of time intervals (in seconds) between publishing of posts for individual user accounts. The median $\bar{\tau}_{uac}^{pst}=43703$ seconds (about 12 hours). The maximum registered interval $\max \tau_{uac}^{pst}=12915073$ (about 149.48 days). There is a considerable proportion of intervals equal to 0, in total – 365349 intervals. Apparently, these 0 intervals are registered, because UnixTime has the resolving power of 1 second.

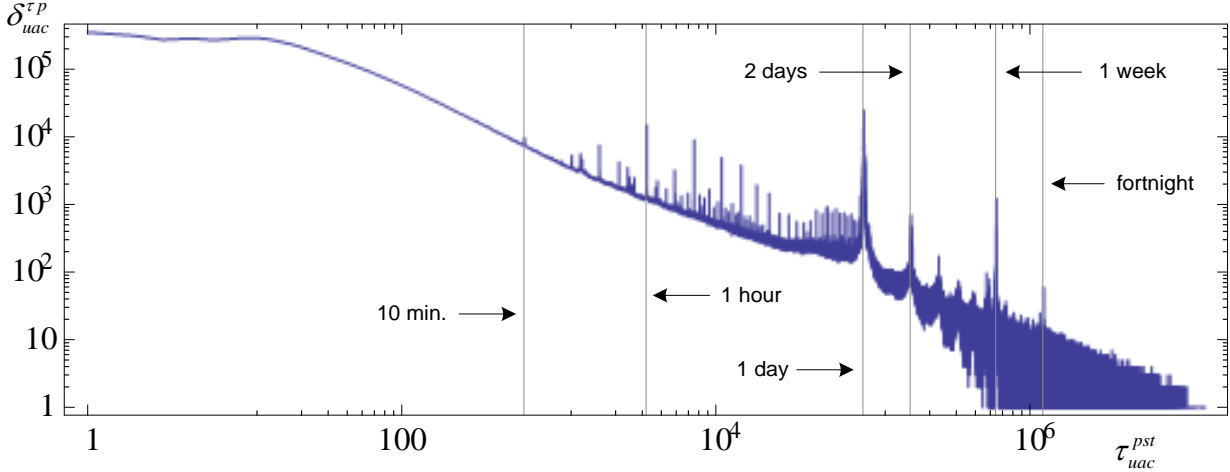

**Fig. 5.** The distribution of time intervals (in seconds) between publishing of posts by individual user accounts.

Figure 5 shows that the most common time intervals between posts of individual user accounts are 1 hour, 1 day, 2 days, 1 week and 2 weeks. The interval of 1 week is much more common than the intervals of 3, 4, 5 days and even than 2 days.

To illustrate the scope of performance values of various accounts, we make the following basic estimates, with normalization to the duration of analysis period:

- 10 posts per user account, about 1 post every two weeks;

- 100 posts per user account, about 2 posts every three days;

- 35 922 posts per user account (the registered maximum), a post every six minutes.

The following points are worth mentioning:

1. The performance of 1 post per week (two weeks) is a typical performance of regular users, writing on topics important to them.

2. The performance of 1-2 posts a day (two days) is a typical performance of professional copywriters or users who actively share photos via their mobile devices.

3. User accounts with higher performance values (over 1 post per hour) are usually communities, news agencies or advertising spam bots.

Summary information on publishing of comments is presented in Table 2.

Summary information on the analyzed period (for comments)

| Parameter | Representation | Value |
|---|---|---|
| Date of publication of the first analyzed comment | $T_b^{cmt}$ | 1 Janurary 2013, 0:0:2 UTC |
| Date of publication of the last analyzed comment | $T_e^{cmt}$ | 2 June 2013, 0:11:9 UTC |
| Duration of the analyzed period | $T_{be}^{pst}$ | ~ 152 days |
| Number of posts with comments | $N^{pst\,c}$ | 5 893 995 |
| Number of authors of posts that have comments. | $N_{pst\,c}^{uac}$ | 660 961 |
| Number of published comments | $N^{cmt}$ | 21 366 037 |
| Number of commentators | $N_{cmt}^{uac}$ | 2 030 855 |

Table 2 can be used to determine basic performance for user accounts by the number of comments:

$$\tilde{S}_{uac}^{cmt} = \frac{N^{cmt}}{N_{cmt}^{uac}} \approx 10.5 \,.$$

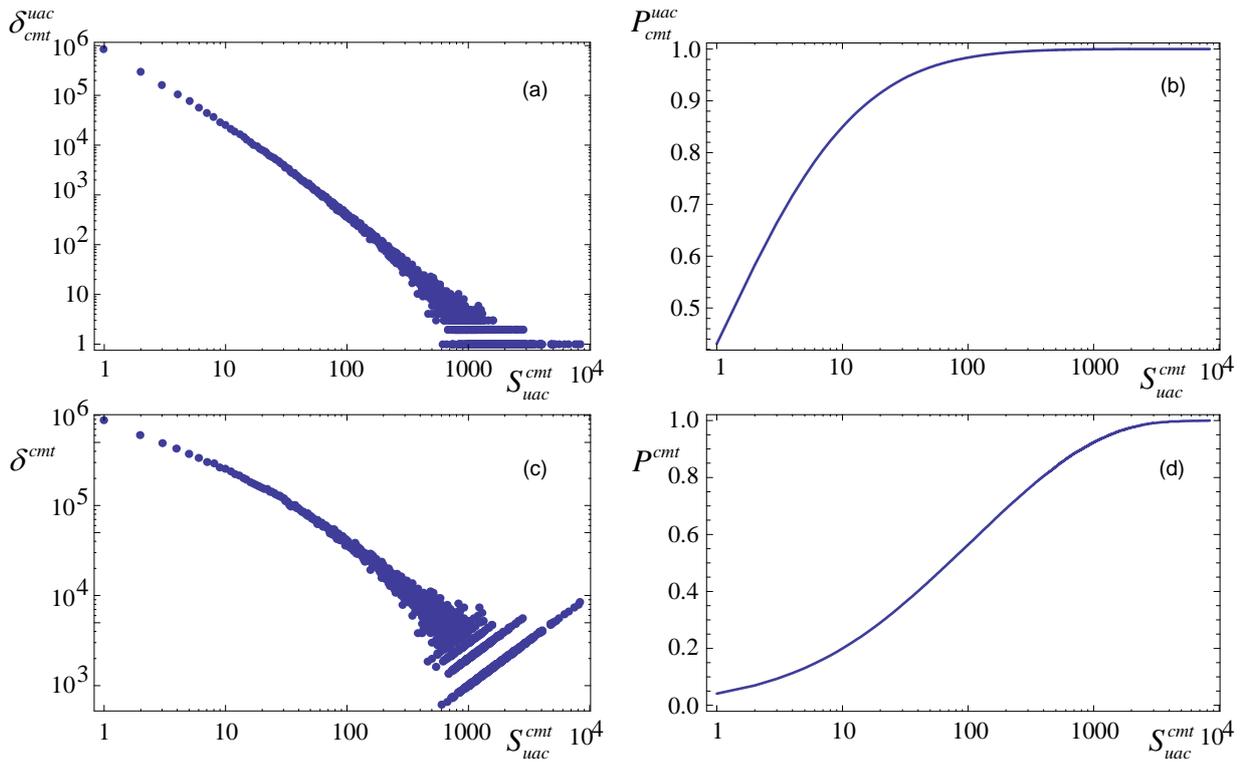

**Fig. 6.** (a) – dependence $\delta_{cmt}^{uac}\left(S_{uac}^{cmt}\right)$; (b) – cumulative share of user accounts, depending on the increase in performance $S_{uac}^{cmt}$; (c) – dependence $\delta^{cmt}\left(S_{uac}^{cmt}\right)$; (d) – cumulative share of published comments, depending on the increase in the user account performance.

In fact, there is a significant variation in performance values (4 orders of magnitude). Figure 6a shows the distribution of the number of user accounts $\delta_{cmt}^{uac}$, depending on the number of published comments $S_{uac}^{cmt}$. The median of $\overline{S}_{uac}^{cmt} \mid P_{cmt}^{uac} = 2$, see Fig. 6b. Maximum performance of max $S_{uac}^{cmt} = 8408$. In Fig. 6c, $\delta^{cmt}$ is the distribution of the number of published comments in the function of user performance $S_{uac}^{cmt}$:

$$\delta^{cmt}\left(S_{uac}^{cmt}\right) = S_{uac}^{cmt}\ \delta_{cmt}^{uac}\left(S_{uac}^{cmt}\right),$$

the median of $\overline{S}_{uac}^{cmt} \mid P^{cmt} = 71$, see Fig. 6d.

Table 2 can be used to make a basic estimate of the number of comments on posts that receive comments:

$$\tilde{S}_{pst\,c}^{cmt} = \frac{N^{cmt}}{N_{cmt}^{pst}} \approx 3.6\,.$$

Figure 7a presents the real distribution $\delta^{pst\,c}$ of the number of posts with comments, depending on the number of comments $S_{pst\,c}^{cmt}$. The median of $\overline{S}_{pst\,c}^{cmt} \mid P^{pst\,c} = 2$, see Fig. 7b. The maximum registered aggregation of max $S_{pst\,c}^{cmt} = 4756$. Figure 7c shows distribution $\delta^{cmt}$ of the number of published comments in the function of the number of comments per post $S_{pst\,c}^{cmt}$:

$$\delta^{cmt}\left(S_{pst\,c}^{cmt}\right) = S_{pst\,c}^{cmt}\ \delta^{pst\,c}\left(S_{pst\,c}^{cmt}\right),$$

the median of $\overline{S}_{pst\,c}^{cmt} \mid P^{cmt} = 7$, see Fig. 7d.

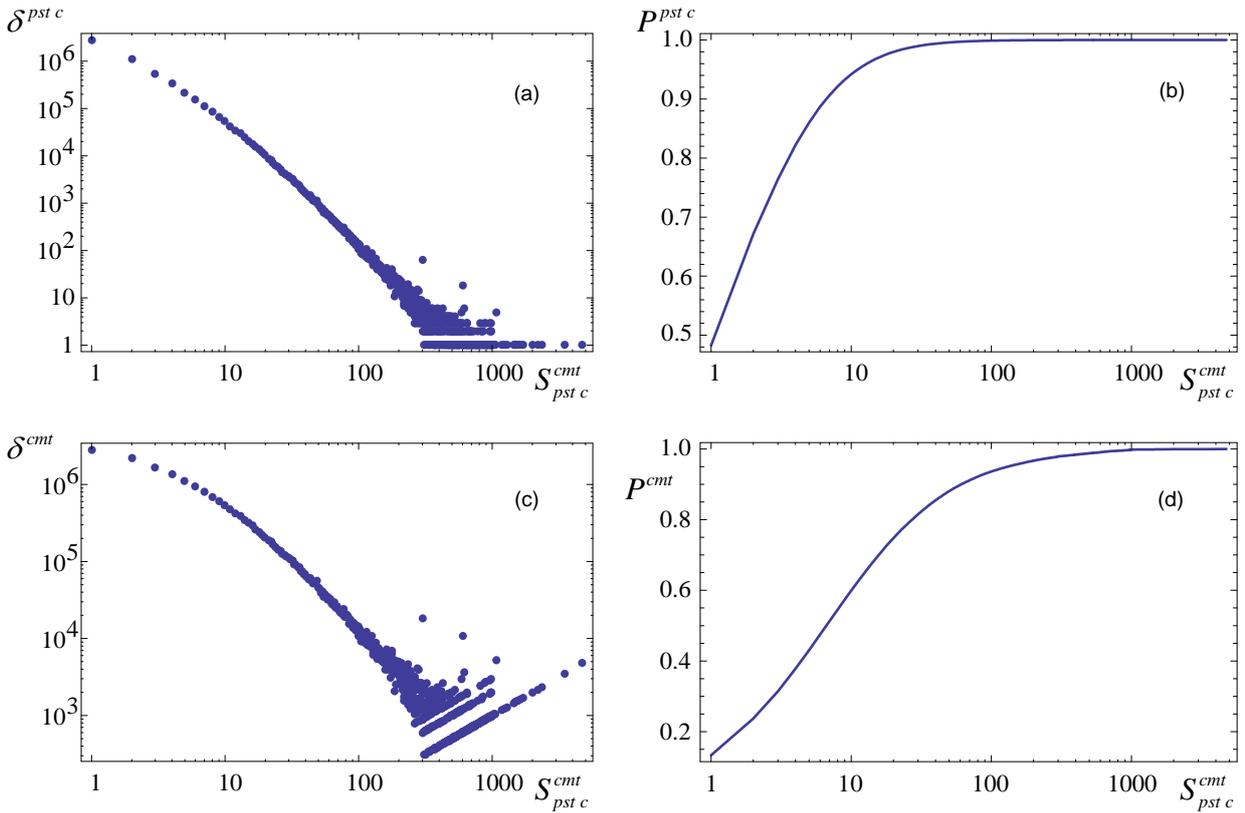

**Fig. 7.** (a) – dependence $\delta^{pst\,c}\left(S_{pst\,c}^{cmt}\right)$; (b) – cumulative share of posts with comments, depending on the increase in the number of comments per post $S_{pst\,c}^{cmt}$; (c) – dependence $\delta^{cmt}\left(S_{pst\,c}^{cmt}\right)$; (d) – cumulative share of published comments, depending on the increase in the number of comments per post $S_{pst\,c}^{cmt}$.

Figure 8a shows distribution $\delta^{pst\,c}$ of the number of posts with comments, depending on the number of comments per post $S_{pst\,c}^{cmt\,s}$ registering only the comments by the author of the post. The median of $\overline{S}_{pst\,c}^{cmt\,s} \mid P^{pst\,c} = 0$, see Fig. 8b. The maximum registered aggregation of $\max S_{pst\,c}^{cmt\,s} = 688$.

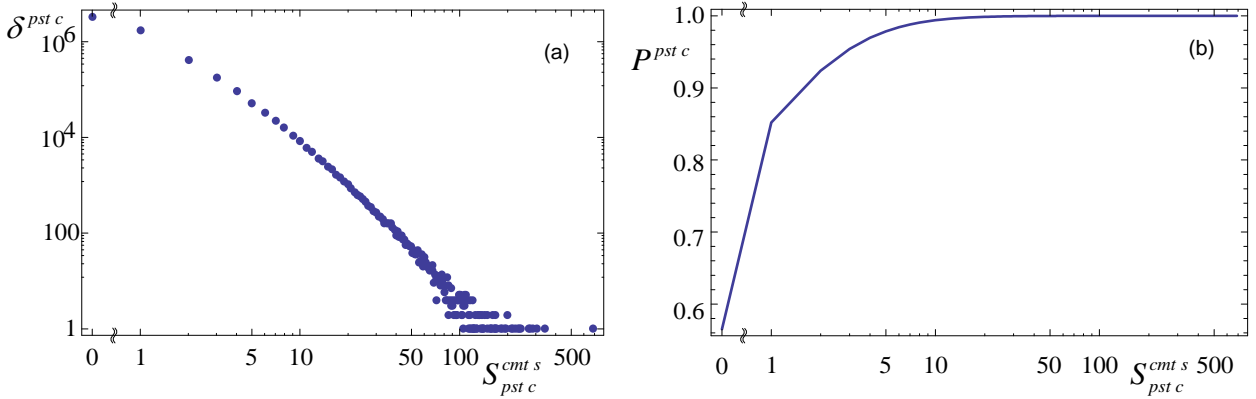

**Fig. 8.** (a) – dependence $\delta^{pst\,c}\left(S_{pst\,c}^{cmt\,s}\right)$; (b) – cumulative share of posts with comments, depending on the increase in the number of comments per post $S_{pst\,c}^{cmt\,s}$ registering only the comments by the author of the post.

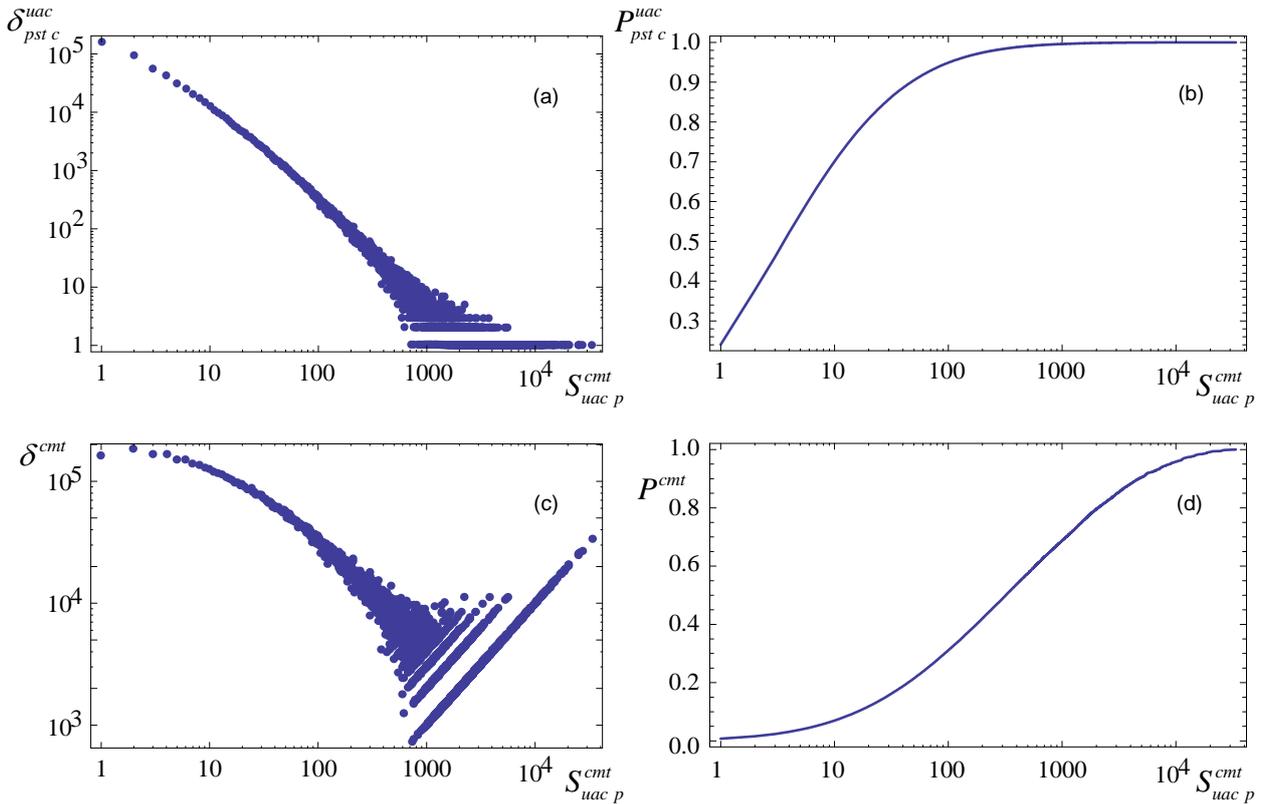

**Fig. 9.** (a) – dependence $\delta_{pst\,c}^{uac}\left(S_{uac\,p}^{cmt}\right)$; (b) – cumulative share of the authors of the posts with comments, depending on the increase in number of comments made by the author of the post $S_{pst\,c}^{cmt}$; (c) – dependence $\delta^{cmt}\left(S_{uac\,p}^{cmt}\right)$; (d) – cumulative share of the published comments, depending on the increase in the number of comments made by the author of the post $S_{pst\,c}^{cmt}$.

Table 2 can be used to make a basic estimate of the number of comments made by the author of the post with comments:

$$\bar{S}_{uac\,p}^{cmt} = \frac{N^{cmt}}{N_{cmt}^{uac}} \approx 32 \,.$$

Figure 9a shows the real distribution $\delta_{pst\,c}^{uac}$ of the number of authors, depending on the number of comments they received $S_{uac\,p}^{cmt}$. The median of $\bar{S}_{uac\,p}^{cmt} \mid P_{pst\,c}^{uac} = 4$, see Fig. 9b. The maximum registered aggregation of $\max S_{uac\,p}^{cmt} = 33585$. Figure 9c shows distribution $\delta^{cmt}$ of the number of published comments in the function of the number of comments $S_{pst\,c}^{cmt}$ registering only the comments by the author of the post:

$$\delta^{cmt}\left(S_{uac\,p}^{cmt}\right) = S_{uac\,p}^{cmt}\; \delta_{pst\,c}^{uac}\left(S_{uac\,p}^{cmt}\right),$$

the median of $\bar{S}_{uac\,p}^{cmt} \mid P^{cmt} = 321$, see Fig. 9d.

Figure 10a shows distribution $\delta^{pst\,c}$ of the number of posts with comments, depending on the number of commentators they attracted $S_{pst\,c}^{uac\,c}$. The median of $\bar{S}_{pst\,c}^{uac\,c} \mid P^{pst\,c} = 1$, see Fig. 10b. The maximum observed aggregation of $\max S_{pst\,c}^{uac\,c} = 2252$.

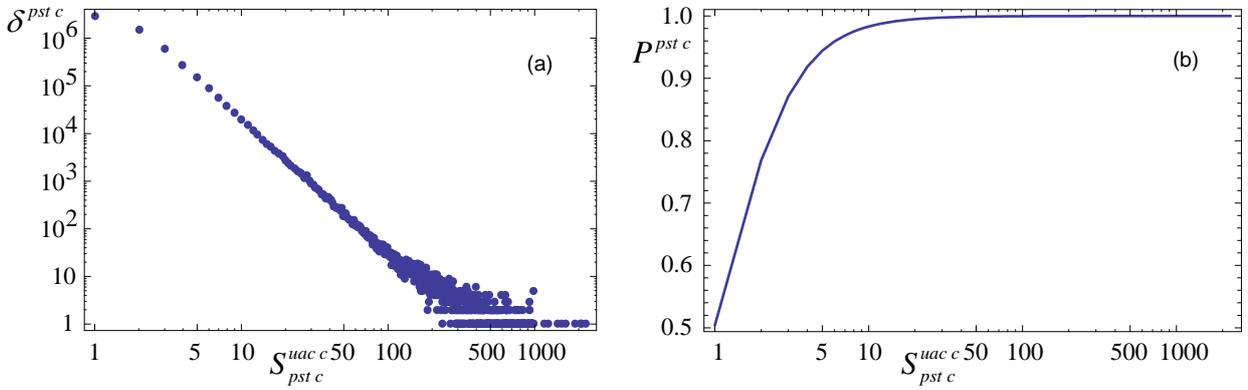

**Fig. 10.** (a) – dependence $\delta^{pst\,c}\left(S_{pst\,c}^{uac\,c}\right)$; (b) – cumulative share of posts with comments, depending on the increase in the number of comments per post.

Figure 11a shows distribution $\delta_{pst\,c}^{uac}$ of the number of authors of posts with comments, depending on the number of commentators they attracted $S_{uac\,c}^{uac\,c}$. The median of $\bar{S}_{uac\,p}^{uac\,c} \mid P_{pst\,c}^{uac} = 3$, see Fig. 11b. The maximum registered aggregation of $\max S_{uac\,p}^{uac\,c} = 12263$.

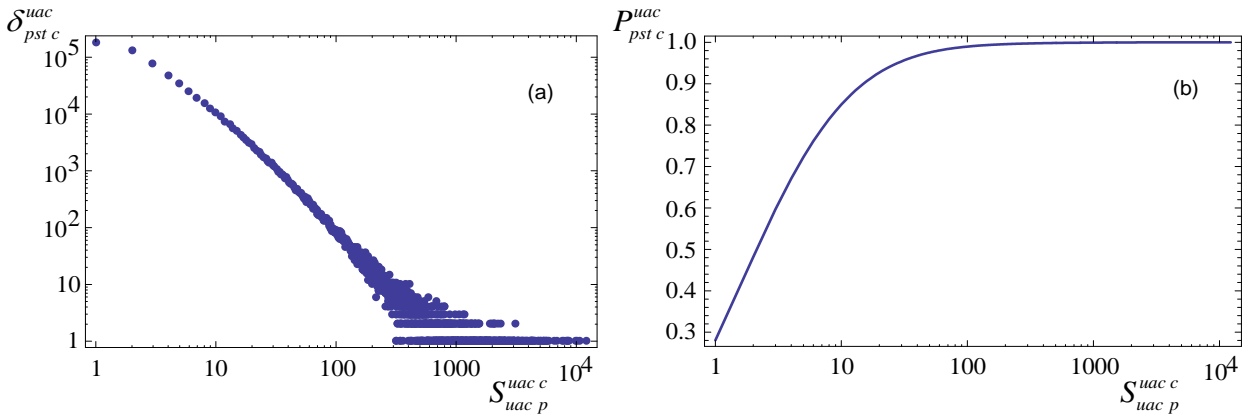

**Fig. 11.** (a) – dependence $\delta_{pst\,c}^{uac}\left(S_{uac\,p}^{uac\,c}\right)$; (b) – cumulative share of authors of posts with comments, depending on the increase in the number of commentators per post.

Figure 12a shows distribution $\delta_{cmt}^{uac}$ of the number of authors of comments, depending on the number of posts per commentator $S_{uac\,c}^{pst\,c}$. The median of $\overline{S}_{uac\,c}^{pst\,c}\,|\,P_{cmt}^{uac}=2$, see Fig. 12b. The maximum registered number of posts per one commentator is $\max S_{uac\,p}^{uac\,c}=4068$.

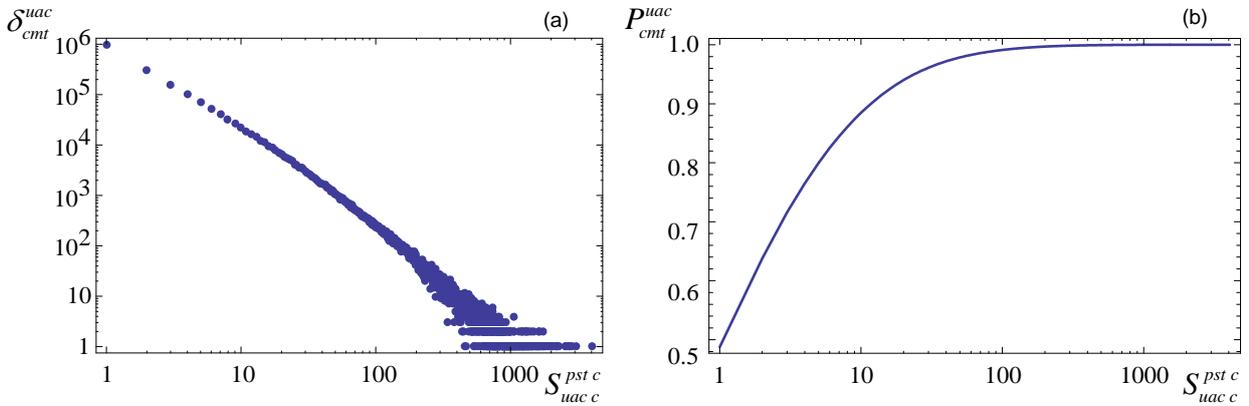

**Fig. 12.** (a) – dependence $\delta_{cmt}^{uac}\left(S_{uac\,c}^{pst\,c}\right)$; (b) – cumulative share of the authors of comments, depending on the increase in the number of posts made by the author of a comment.

Figure 13a shows distribution $\delta_{cmt}^{uac}$ of the number of authors of comments, depending on the number of authors of posts per commentator $S_{uac\,c}^{uac\,p}$. The median of $\overline{S}_{uac\,c}^{uac\,p}\,|\,P_{cmt}^{uac}=1$, see Fig. 13b. The maximum registered number of authors of posts per commentator is $\max S_{uac\,c}^{uac\,p}=521$.

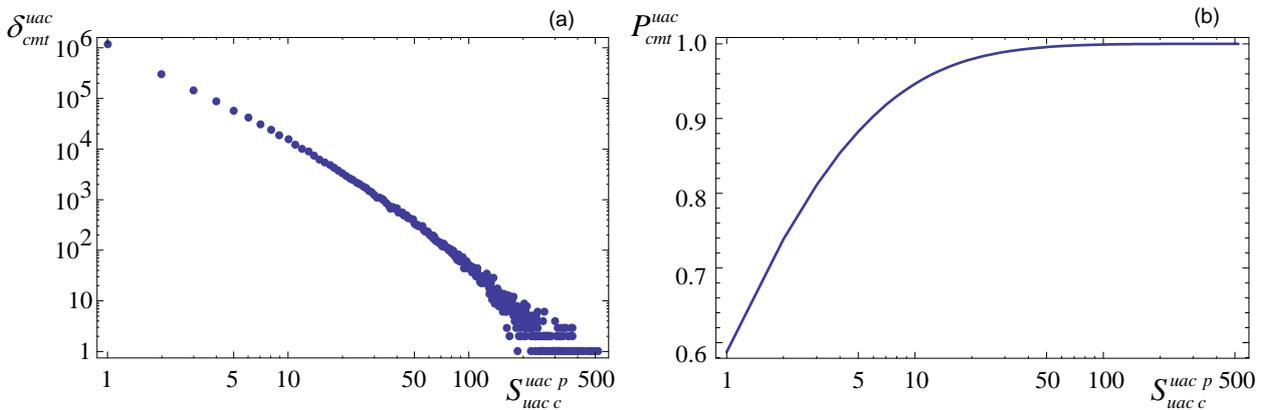

**Fig. 13.** (a) – dependence $\delta_{cmt}^{uac}\left(S_{uac\,c}^{uac\,p}\right)$; (b) – cumulative share of authors of comments, depending on the increase in the number of authors of posts per author of a comment.

Figure 14a shows distribution $\delta^{pst\,c}$ of the number of posts with comments, depending on the time delay from the moment of publication of the post to the publication of the first comment $\tau_{f\,c}^{pst}$. The median $\overline{\tau}_{f\,c}^{pst}\,|\,P^{pst\,c}=3076$ seconds (about 51 minutes). The maximum registered time delay $\max \tau_{f\,c}^{pst}=182274110$ seconds (about 5 years 9 months), see Fig. 14b. The most common time delay is 50 seconds (see area g50 in Fig. 14a). Time delay intervals of $0-86$ seconds have been registered. Apparently, these 0 intervals are registered, because UnixTime has the resolving power of 1 second.

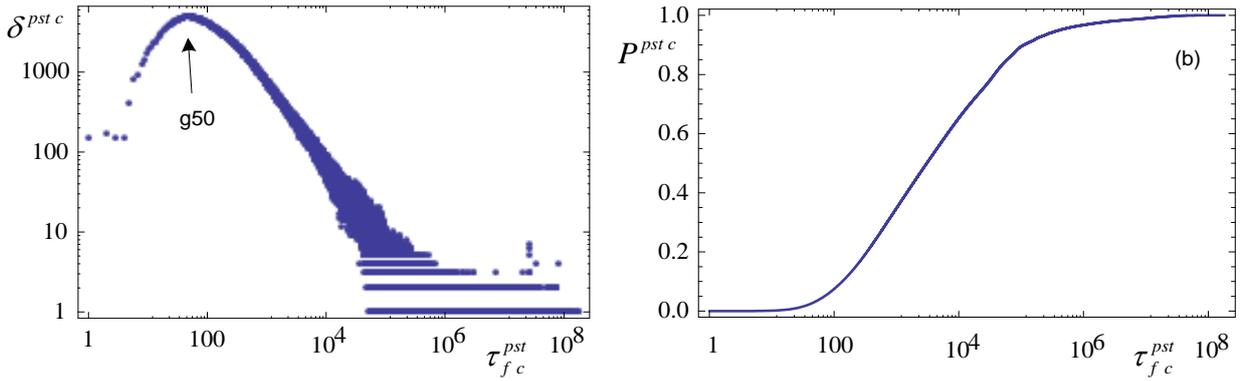

**Fig. 14.** (a) – dependence $\delta^{pst\,c}\left(\tau_{f\,c}^{pst}\right)$; (b) – cumulative share of posts with comments, depending on the increase in the delay between the moment of publication of the post and the time of publication of the first comment.

It stands to mention that negative values of $\tau_{f\,c}^{pst}$ have been registered, 18273 in total or 0.31 % of all the posts with comments. The median for negative time values $\overline{\tau}_{f\,c}^{pst}\,|\,P^{pst\,c}:\tau_{f\,c}^{pst}<0=-180523$ seconds (about 2 days 2 hours), see Fig. 15. The maximum registered negative delay time $\min\tau_{f\,c}^{pst}=-11901711$ seconds, whereby the publishing of a comment precedes the publishing of the post by about 4 months 17 days. A bug in Facebook can be one of the possible causes [7].

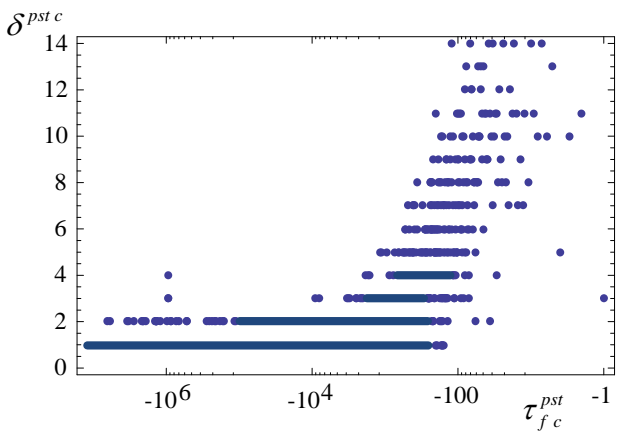

**Fig. 15.** Dependence $\delta^{pst\,c}\left(\tau_{f\,c}^{pst}\right)$ for negative delay time values between the moment of publication of the post and the time of publication of the first comment.

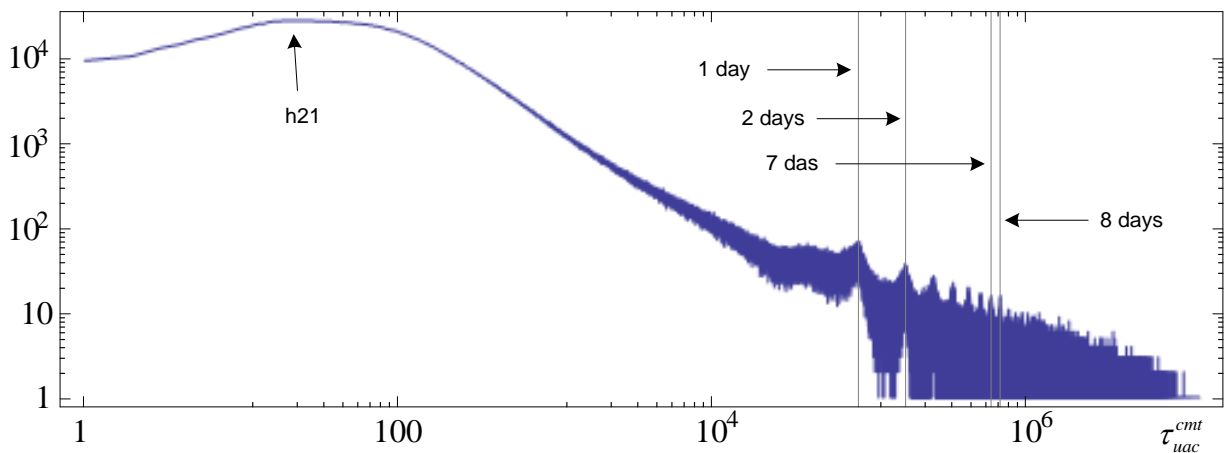

**Fig. 16.** The distribution of time intervals (in seconds) between publishing of comments by different user accounts.

Distribution $\tau_{uac}^{cmt}$ of the time intervals (in seconds) between the publishing of comments by different user accounts is shown in Fig. 18. The median $\overline{\tau}_{uac}^{cmt} = 5163$ seconds (about 86 minutes). The maximum registered interval $\max \tau_{uac}^{cmt} = 12900451$ (about 149.311 days). There is a considerable number of 0 intervals, 3265613 in total. Apparently, these 0 intervals are registered, because UnixTime has the resolving power of 1 second.

Figure 16 shows that the most common intervals are from 1 to 8 days with 24 hour increments. There is a notable local maximum of 21 seconds (h21).

## 4. Conclusion

The article presents the results of the analysis of Facebook, which has been carried out from the standpoint of the dynamics of publishing of posts and comments. The object of the study is the Russian segment of the network in the period from 1.01.2013 to 02.06.2013. The content of posts and comments has not been analyzed. Two initial parameters have been used in the analysis: author and time of publication of a message.

The analysis and the calculations have yielded a set of descriptive values that provide insight into the processes of publishing of posts and comments by the users of Facebook. Several distinguished features have been identified:

1) A significant anomaly in the number of user accounts with the performance of approximately two posts per three days has been detected. The total number of such accounts exceeds the theoretical value by $96623_{-5181}^{+0}$. The theoretical value has been calculated by approximated power distribution.

2) About 50 % of all posts are published by users with the performance of no more than three posts per two days.

3) The average time interval (median) between publishing of posts for individual user accounts is about 12 hours.

4) About 50% of all comments are published by users with the performance of no more than one comment per two days.

5) About 50% of the posts that have at least one comment have no more than two comments.

6) About 50% of all comments are made on posts that receive no more than 7 comments.

7) About 55% of posts with at least one comment are not commented by the authors themselves.

8) About 50 % of users with posts that receive at least one comment, in total, get no more than 4 comments each (during the period of the analysis ~ 152 days).

9) About 50 % of all comments are received by authors of posts (with at least one comment) that, in total, receive no more than 321 comments each (during the period of the analysis).

10) About 50% of the posts that have at least one comment have no more than one commentator.

11) About 50 % of all comments are received by authors of posts (with at least one comment) that have no more than three commentators.

12) About 50% of all commentators commented on no more than one post each (during the period of the analysis ~ 152 days).

13) About 50% of all commentators commented on posts of no more than one other user account each (during the period of the analysis ~ 152 days).

14) About 50% of the posts receive the first comment within 51 minutes after the publication of the post. The most probable time of first comment is 50 seconds after the publication of the post.

The maximum registered delay time between the publication of the post and the first comment is 5 years 9 months.

15) We have determined that 18273 posts have negative time difference between their publication moment and the first comment [7]. About 50% of these posts have negative delay time of up to 20 days 2 hours.

16) About 50% of the commentators (with two or more subsequent comments) publish comments at the interval of no more than 86 minutes. This interval have a notable local maximum of 21 seconds.

Our results allow to assess the dynamics of publishing of posts and comments in the Russian segment of Facebook. These results are important for addressing a number of other issues, including:

- Studying the structural (topological and metric) characteristics of network-describing links between commentators and posts authors.

- Identifying sustainable communities of users from the standpoint of the stationary nature of their relations when commenting on posts.

- Identifying the most efficient user accounts using the criterion of their information activity (their appeal to the community of users).

- Developing an adequate quantitative model for studying and forecasting informational activity of Facebook users (in terms of writing posts and comments).

## Acknowledgements

The author is grateful to Professor D. A. Novikov for his support and interest to this study, as well as to Ph.D. D. A. Gubanova for the preparation of the initial data.

## References

1. Ahlqvist T., Bäck, A., Halonen M. and Heinonen S. *Social media road maps exploring the futures triggered by social media //* VTT Research Notes (2008) **2454**:13.

2. Kaplan A.M. and Haenlein M. *Users of the world, unite! The challenges and opportunities of social media. //* Business Horizons (2010) **53**:1. p. 61.

3. Governor J., Hinchcliffe D. and Nickull D. *Web 2.0 Architectures: What Entrepreneurs and Information Architects Need to Know.* O'Reilly (2009).

4. Hassanien A.E., Abraham A. and Snšel V. *Computational Social Network Analysis Trends, Tools and Research Advances.* New York: Springer (2009).

5. Bing L. *Web Data Mining: Exploring Hyperlinks, Contents, and Usage Data.* New York: Springer (2011).

6. *Resource Description Framework (RDF) Model and Syntax Specification.* Eds: O. Lassila and Ralph R. Swick // W3C Proposed Recommendation 05 January 1999.

7. FaceBook Developers: Bugs created_time incorrect in Posts (2013, Nov.). [Online]. Available: URL: https://developers.facebook.com/bugs/514380435304761

**Andrey V. Makarenko** – was born in 1977, since 2002 -- Ph. D. of Cybernetics. Founder and leader of the Research & Development group "Constructive Cybernetics". Author and coauthor of more than 60 scientific articles and reports. Member IEEE (IEEE Signal Processing Society Membership; IEEE Computational Intelligence Society Membership). Research interests: Analysis of the structure dynamic processes, predictability; Detection, classification and diagnosis is not fully observed objects (patterns); Synchronization and self-organization in nonlinear and chaotic systems; System analysis and math. modeling of economic, financial, social and bio-physical systems and processes; Convergence of Data Science, Nonlinear Dynamics and Network-Centric.